\documentclass[letterpaper, 10 pt, conference]{ieeeconf}  

\IEEEoverridecommandlockouts                              
\usepackage{cite}
\usepackage{amsmath,amssymb,amsfonts}
\usepackage{algorithmic}
\usepackage{graphicx}
\usepackage{textcomp}
\usepackage{xcolor}
\usepackage{tikz}
\usetikzlibrary{shapes,arrows,positioning,calc}
\usepackage{caption}
\usepackage{subcaption}
\newtheorem{remark}{Remark}[section]

\title{\LARGE \bf
A robust but easily implementable remote control  for quadrotors: \\ Experimental acrobatic flight tests
}

\author{Maison Clouatre$^{\ast }$, Makhin Thitsa$^{\ast}$, Michel Fliess$^{+ \dag}$, C\'{e}dric Join$^{\P \dag}$\\
$^{\ast}$Department of Electrical and Computer Engineering, Mercer University \\ Macon GA 31207, USA  \\maison.roland.clouatre@live.mercer.edu, thitsa\textunderscore m@mercer.edu\\
$^{+}$LIX (CNRS, UMR 7161), \'Ecole polytechnique \\ 91128 Palaiseau, France \\Michel.Fliess@polytechnique.edu  \\
$^{\P}$  CRAN (CNRS, UMR 7039)), Universit\'{e} de Lorraine \\ BP 239, 54506 Vand{\oe}uvre-l\`{e}s-Nancy, France \\
cedric.join@univ-lorraine.fr \\
$^{\dag}$ AL.I.E.N., 7 rue Maurice Barr\`{e}s, 54330 V\'{e}zelise, France \\
\{michel.fliess, cedric.join\}@alien-sas.com
}

\begin{document}
\tikzset{
block/.style = {draw, fill=white, rectangle, minimum height=3em, minimum width=3em},
tallblock/.style = {draw, fill=white, rectangle, minimum height=11em, minimum width=3em},
tmp/.style  = {coordinate},
sum/.style= {draw, fill=white, circle, node distance=1cm},
input/.style = {coordinate},
output/.style= {coordinate},
pinstyle/.style = {pin edge={to-,thin,black}
}
}

\maketitle
\thispagestyle{empty}
\pagestyle{empty}

\begin{abstract}
Experimental flight tests are reported about quadrotors UAVs via a recent model-free control (MFC) strategy, which is easily implementable. We show that it is possible to achieve acrobatic rate control of the UAV, which is beyond the previous standard. The same remote controller is tested on two physical vehicles without any re-tuning. It produces in both cases low tracking error. We show that MFC is robust even when the quadrotor is highly damaged. A video footage can be found at: {\tt https://youtu.be/wtSLalA4szc}


\end{abstract}


\section{Introduction}
Quadrotors are today the most ubiquitous unmanned aerial vehicles (UAVs) (see, \textit{e.g.}, \cite{stevens2015aircraft,quan,hassa}, and the references therein). Many remote control strategies, including machine learning techniques, have been investigated (see, \textit{e.g.}, \cite{stevens2015aircraft,quan,hassa,zulu,pounds,lambert}, and the references therein). According to some authors (see, \textit{e.g.}, \cite{pounds}, and the references therein) the simplicity of PIDs explains why they behave better in practice than more advanced controllers. Nevertheless the well-known shortcomings of PIDs (see, \textit{e.g.}, \cite{astrom,murray}) remain valid in this UAV context. This is why we suggest here \emph{model-free control} (MFC) in the sense of \cite{fliess2013model} which might be viewed as an improvement of PIDs. MFC, which has been successfully applied to a number of concrete case-studies (see \cite{fliess2013model,bara,ftc} and the references therein for most publications until the beginning of 2020), has already been employed several times for UAVs \cite{al2016robust,wang,bekcheva,barth1,barth2,chek,ozbek}.  This communication might be the first one about experimental tests of MFC with quadrotors.

We are concerned here with the \emph{acrobatic} flight\footnote{There is a legal definition for \emph{acrobatic}, or \emph{aerobatic}, flights: \newline {\tt https://www.law.cornell.edu/cfr/text/14/91.303} \newline See, \textit{e.g.}, \cite{aerobat,acrobat} for brillant and recent contributions.}  of quadrotors. This type of flight is important to tracking, evasion, and rescue missions, as well as obstacle avoidance in dense environments. We propose a fully MFC-based flight controller for robust rate tracking onboard quadrotor UAVs, which is an improvement beyond the previous standard. The proposed method is validated
\begin{itemize}
\item on-board various physical UAVs,
\item via highly damaged vehicles.
\end{itemize}
In both cases
\begin{itemize}
\item no modification or re-tuning of the controller is needed,
\item the UAVs are always well stabilized around the reference trajectory.
\end{itemize}
It highlights the method's robustness and data-driven nature.
MFC requires little computational power to implement, thus it is an exciting low-cost alternative to PID that is used on most quadrotors today.

The paper is organized as follows. In Section \ref{Sec:MFC} a short review of MFC theory is provided. Section \ref{Sec:Implementation} outlines the implementation and deployment of MFC on-board quadrotor UAVs. Section \ref{Sec:Exp} reports experimental flight data for various vehicles under MFC and compares the flight performance achieved to that of a PID, the most common control technique used on small UAVs. Conclusions and future works follow.

\section{Model-free control: A short review}
\label{Sec:MFC}
\subsection{The ultra-local model for a SISO systelm}
It has been proved in \cite{fliess2013model} that under quite weak assumptions any SISO system, with input $u$ and output $y$, may be well approximated by the \emph{ultra-local} model
\begin{equation}
    y^{(m)} = F(t) + \alpha u
    \label{eq:ModelFree}
\end{equation}
where
\begin{itemize}
\item $m \geq 1$ is the derivation order,
\item the time-varying quantity $F$ subsumes not only the un-modeled dynamics, but also the external disturbances,
\item the constant $\alpha \in {\mathbb R}$ is such that the three quantities $y^{(m)}$, $F$, $\alpha u$ in  Eq. \eqref{eq:ModelFree} are of the same order of magnitude.
\end{itemize}
Note that
\begin{itemize}
\item the poorly known plant is  not necessarily of order $m$: $y^{(\nu)}$, where $\nu > m$ may be sitting in $F$,
\item in almost all the numerous concrete case-studies that were encountered until now, $m = 1$, with only some exceptions where $m = 2$,
 \item it is meaningless to try to estimate $\alpha$ precisely.
\end{itemize}

\subsection{MIMO systems}
\label{mimo}
Consider a multi-input multi-output (MIMO) system with $p$ control variables $u_i$ and $p$ output variables $y_i$, $i = 1, \dots, p$. It has been observed in \cite{toulon} and confirmed by all encountered concrete case-studies (see, \textit{e.g.}, \cite{wang0}), that such a system may be regulated via $p$ monovariable ultra-local models:
\begin{equation*}\label{multi}
y_{i}^{(m_i)} = F_i + \alpha_i u_i
\end{equation*}
where $F_i$ may also depend on $u_j$, $y_j$, and their derivatives, $j = 1, \dots, p$.
\begin{remark}
In our example $p = 3$, one with a first (resp. second) order ultra-local model. Previous publications on UAVs \cite{wang,bekcheva,barth1} teach us that $m = 2$ is often appropriate (see also \cite{miras,menhour,tich,haddar}) for other examples).\footnote{According to \cite{fliess2013model}, the choice $m = 2$ is necessary if there is (almost) no friction. The question however is far from being fully understood \cite{iste}.}
\end{remark}

\subsection{iP and iPD}
\subsubsection{$m = 2$}\label{m=2}
Eq. \eqref{eq:ModelFree} becomes
\begin{equation}
    \ddot{y} = F(t) + \alpha u.
    \label{2}
\end{equation}
Associate \cite{fliess2013model} to Eq. \eqref{2} the \emph{intelligent Proportional-Derivative} controller, or \emph{iPD},
\begin{equation}\label{ipd}
u = - \frac{\hat{F} - \ddot{y}_r + K_P e + K_D \dot{e}}{\alpha}
\end{equation}
where
\begin{itemize}
\item $\hat{F}$ is an estimate of $F$,
\item $y_r$ is the reference trajectory,
\item $e = y - y_r$ is the tracking error,
\item $K_P, K_D \in {\mathbb R}$ are the feedback gains
\end{itemize}
It yields
$$
\ddot{e} - K_D \dot{e} - K_I e = F - \hat{F}.
$$
If the estimate $\hat{F}$ is ``good'', \textit{i.e.}, $F - \hat{F} \approx 0$, the choice of the gain $K_P$, $K_D$ for ensuring ``local'' stability around the reference trajectory is straightforward. This is a major difference with classic PIs and PIDs (see, \textit{e.g.}, \cite{astrom,murray}). Moreover and obviously no anti-windup is needed for iPDs.
\begin{remark}\label{sensor}
Here $\dot{e}$  in Eq. \eqref{ipd} is given by an  appropriate sensor for measuring $\dot{y}$.
\end{remark}
\subsubsection{$m = 1$} Eq.  \eqref{eq:ModelFree} becomes
\begin{equation*}
    \dot{y} = F(t) + \alpha u
    \label{1}
\end{equation*}
The corresponding \cite{fliess2013model}  \emph{intelligent Proportional} controller, or \emph{iP}, reads
\begin{equation}
u = - \frac{\hat{F} - \dot{y}_r + K_P e}{\alpha}.
\label{ip}
\end{equation}
The adaptation of Section \ref{m=2} is straightforward.

\subsection{Estimation of $F$}\label{estim}
A classic result from mathematical analysis (see, \textit{e.g.}, \cite{rudin}) states that under a weak integrability condition any function $\Phi: [a, b] \rightarrow \mathbb{R}$, $a, b \in \mathbb{R}$, $a < b$, may be approximated by a \emph{step} function, \textit{i.e.}, a piecewise constant function. Then, according to \cite{fliess2013model}, an estimate $\hat{F}$ of $F$ is computed by averaging  $F(t) = y^{(m)}(t) - \alpha u(t)$, which is deduced from Eq. \eqref{eq:ModelFree}, on a ``short'' sliding time window. If $m = 1$, Remark \ref{sensor} tells us that $\dot{y}$ is obtained via a sensor. If $m = 2$, set $\ddot{y}(t) \approx \frac{\dot{y}(t) - \dot{y}(t - T)}{T}$, where $T > 0$ is the sampling period. A \emph{Finite Impulse Response}, or \emph{FIR}, filter is of course used in practice.
\begin{remark}
The unavoidable noise corruptions are attenuated via the averaging integral (see \cite{bruit} for a mathematical explanation, and \cite{sira,morales} for applications to parameter identification and signal processing).
\end{remark}

\section{Quadrotor Dynamics \& Control}
\label{Sec:Implementation}
Many authors have reported models of quadrotor dynamics \cite{stevens2015aircraft,quan}. We refer interested readers to said models to build some intuition about the behavior of such a vehicle and to recognize their nonlinearities, which are potent during acrobatic flight. Naturally, these are only approximate models of the quadrotor's dynamics; therefore, a controller that builds its own model from data is highly desirable.

\subsection{Quadrotor Attitude}
A diagram of a quadrotor is shown in Figure \ref{fig:Quad} and is useful when considering its behavior. We consider the case of a 3 degree-of-freedom (DOF) quadrotor with controllable roll ($\phi$), pitch ($\theta$), and yaw ($\psi$) axes. The UAV's rate of rotation, or attitude, about each of these axes---$\dot{\phi}, \dot{\theta},$ and $\dot{\psi}$---is to be controlled by automatically adjusting the speed of each of the vehicle's four propellers, $\omega_1-\omega_4$. The exact relationship between the vehicle's attitude and these speeds must be inferred by the controller. Some intuition may be gained from this figure. For instance, if one wanted to roll to the right, they would increase the speed of motor 4 and decrease the speed of motor 2. Pitching forward would entail increasing the speed of motor 3 and decreasing that of motor 1. Finally, yawing counter-clockwise would require increasing the speed of the clockwise rotating motors and decreasing the speed of the counter-clockwise motors. This intuition is used in Section \ref{sec:ControlScheme} in order to deliver control signals to the motors. Only this low-level knowledge is given to the controller. All other unknown dynamics must be accounted for numerically.

\subsection{Quadrotor Control Scheme}
\label{sec:ControlScheme}

\begin{table}[t]
\caption{Control Parameters}
\label{table:1}
\centering
\begin{tabular}{|c|| c c c c c|}
 \hline
 \textbf{Parameter:} & $m$ & $\alpha$ & $T$  & $K_d$ & $K_p$ \\ [0.5ex]
 \hline\hline
 \textbf{Roll:} & 2 & 1 & 0.02s & 0.096 & 3.0\\
 \textbf{Pitch:} & 2 & 1 & 0.02s & 0.096 & 2.7\\
 \textbf{Yaw:} & 1 & 1 & 0.02s &  & 2.7\\[1ex]
 \hline
\end{tabular}
\end{table}

A single instantiation of MFC will be assigned to each of the quadrotor's degrees of freedom: roll, pitch, and yaw. Thus, a multi-variable control formulation will be implemented as outlined in Section \ref{mimo}. Three control inputs--- $u_\phi, u_\theta, \text{ and } u_\psi$---will be used to manipulate the vehicle's attitude. The output of these monovariable controllers must be translated into four separate control signals for each of the vehicle's four motors. The appropriate control signals for motors 1--4 are as follows:
\begin{equation}
    \boldsymbol{u}_{\text{motors}}=
    \begin{bmatrix}
    u_1 \\
    u_2 \\
    u_3 \\
    u_4
    \end{bmatrix}
    =
    \begin{bmatrix}
    u_t - u_\theta - u_\psi\\
    u_t - u_\phi + u_\psi\\
    u_t + u_\theta - u_\psi\\
    u_t +u_\phi + u_\psi
    \end{bmatrix}.
    \label{eq:motors}
\end{equation}
Here, $u_t$ is the baseline throttle given to the quadrotor by its pilot. If the vehicle were stationary and undisturbed, $u_t$ would be used by the pilot to adjust the height of the quad. A block diagram visualization of the proposed control scheme is shown in Figure \ref{fig:Scheme}.

In our physical implementation, iPDs are used to control the pitch and roll axes. Due to the high level of drag associated with the yaw axis, an iP is more appropriate. In traditional linear control of quadrotors, it is common to use PID on the pitch and roll axes while employing a PI controller on the yaw axis. It has been shown that the appropriate replacement to a PI is an iP and a the appropriate replacement to a PID is an iPD \cite{fliess2013model}. Due to the decoupled nature of control variables, we can use ultra-local models of varying order without fear of cross-talk or adverse effects.

The control parameters used in our physical implementation are listed in Table \ref{table:1}. In each installment of MFC, elementary numerical methods are employed for the implementation (see Sect. \ref{estim}). The output of each of the three controllers is appropriately saturated so that the commands sent to each of the four motors is reasonable. Rate measurements taken by a MPU6050 gyroscope are passed through a simple complimentary filter. The output of the filter is a convex combination of the current measurement and the previous filter output. The controller must overcome the inherent noise that is allowed through this low-level filter. Thus, quality results will support the notion of robustness of the proposed technique. The controller is implemented on an 8-bit Atmega 328 micro-controller and uses a refresh rate of 250 Hz. The inexpensive hardware used here demonstrates that this control technique can be implemented with relatively little computational power (see also \cite{hardware})---making it a very attractive nonlinear alternative to traditional PID used profusely on small UAVs.

\section{Experimental Results\protect\footnote{A video is available at: {\tt https://youtu.be/wtSLalA4szc}}}
\label{Sec:Exp}
We demonstrate the robustness of the proposed technique by using the same control scheme across vehicles with varying dynamics. The controller is not re-tuned between any of these experiments. All flights were flown outdoors on the campus of Mercer University in Macon, GA, USA, on the 18$^\text{th}$ of March, 2020. On this day there were southern winds blowing upwards of 9 mph.

\subsection{DJI F450 Quadrotor}
\begin{figure*}
    \centering
    \includegraphics[scale=0.95]{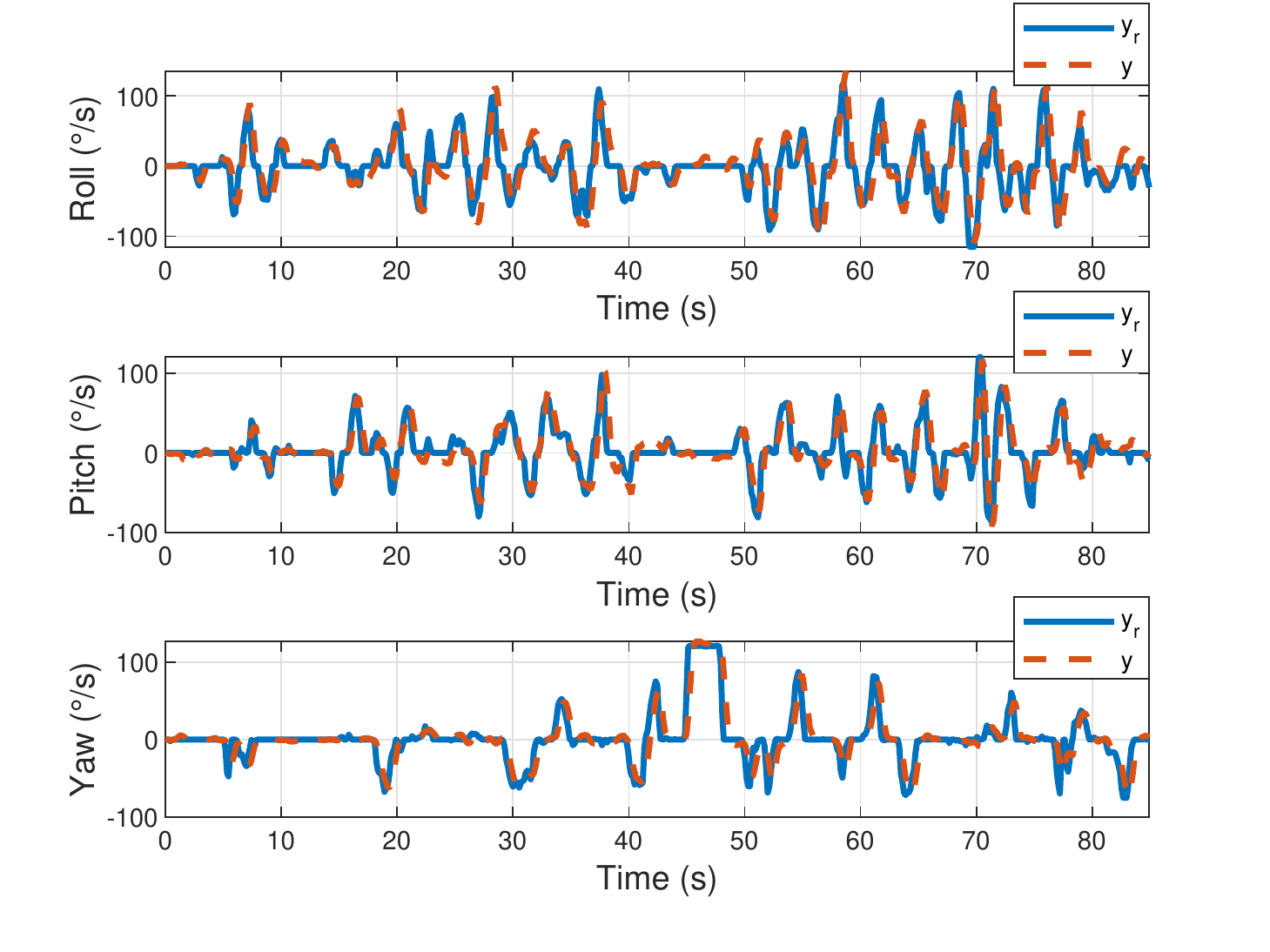}
    \caption{DJI F450 roll, pitch, and yaw performance.}
    \label{fig:F450}
\end{figure*}

The first vehicle flown was a DJI F450 quadrotor equipped with readily available brushless motors and affordable plastic propellers. It is a light-weight drone whose center of gravity is located at the center of the body of the vehicle. The goal of the controller is to minimize the error between rate commands provided by the UAV's pilot and the rates at which the vehicle is rotating. The pilot flew the drone around an open field performing various maneuvers, \textit{i.e.}, the reference trajectories are arbitrary and are generated in real time. The performance of the drone is shown in Figure \ref{fig:F450}. The mean absolute error for each degree of freedom was under 10 degrees per second, which is remarkable since no high-level filtering was used on sensor measurements.

\subsection{DJI F450 Minus Half Its Propellers}
\begin{figure*}
    \centering
    \includegraphics[scale=0.95]{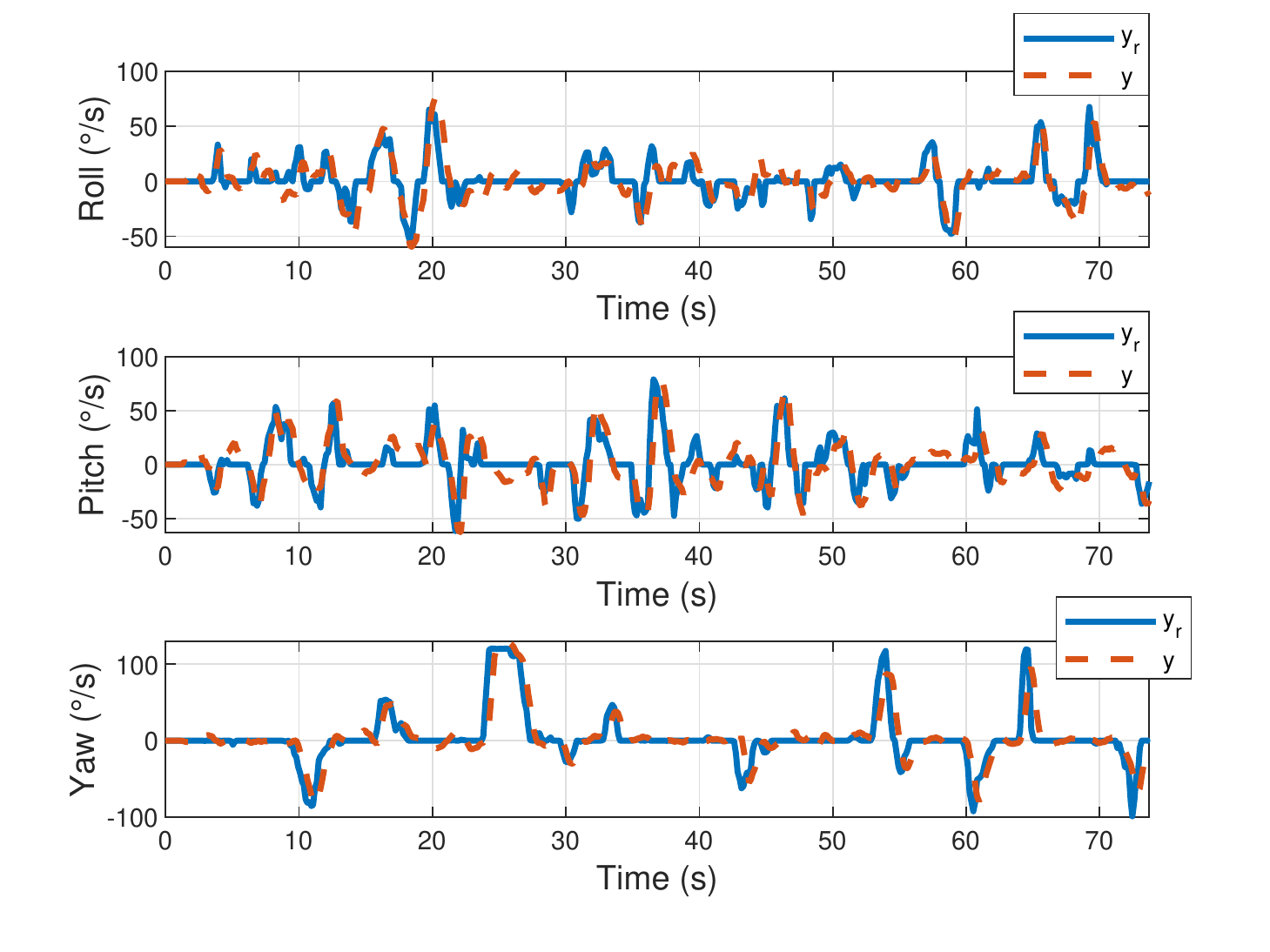}
    \caption{DJI F450 roll, pitch, and yaw performance with severely damaged propellers.}
    \label{fig:NoProps}
\end{figure*}

\begin{figure*}
    \centering
    \includegraphics[scale=0.18]{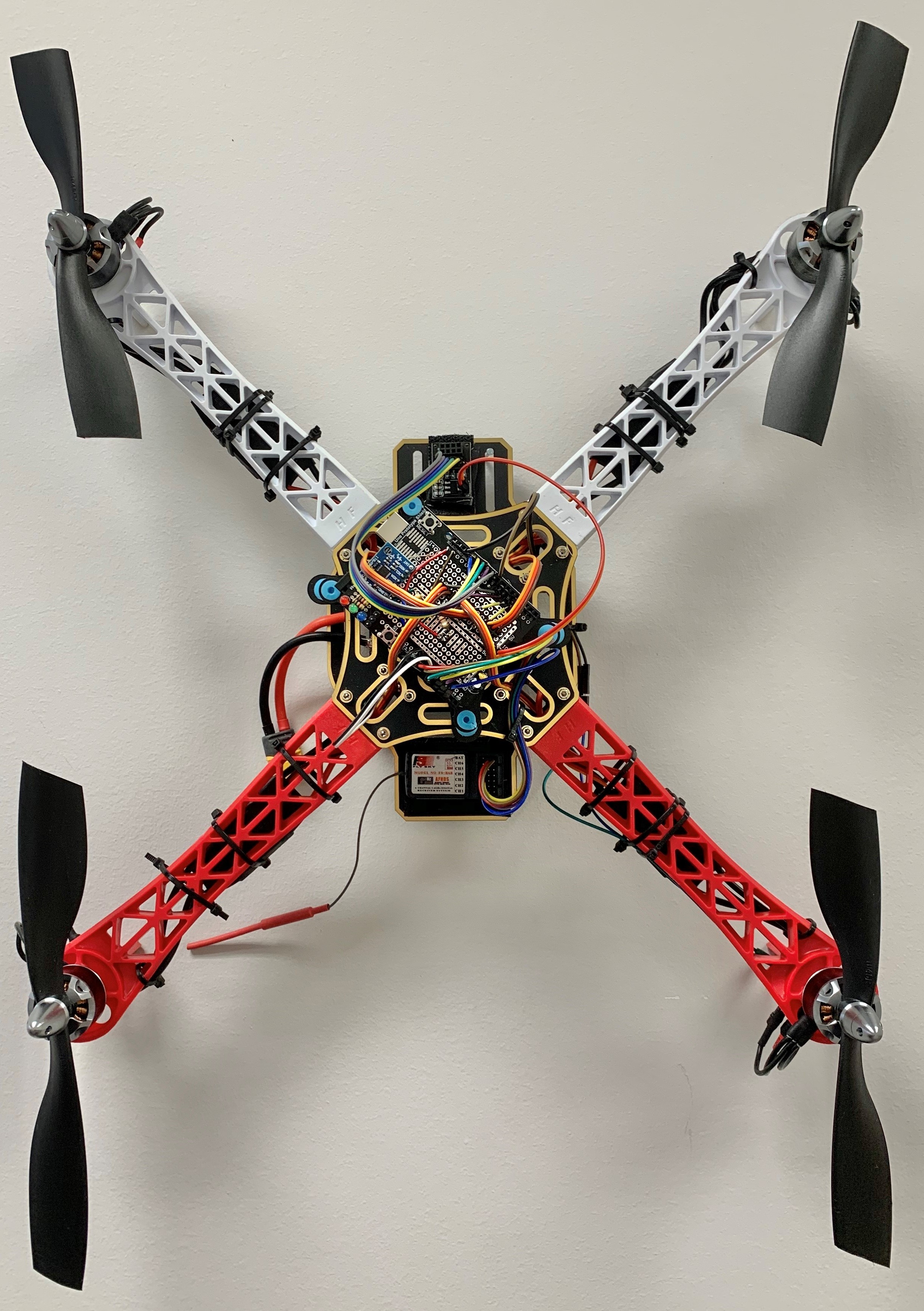}
    \caption{A DJI F450 with damaged rotors. This serves to simulate a quadrotor being damaged in flight.}
    \label{fig:Damaged}
\end{figure*}

The dynamics of the quadrotor were then changed by cutting off significant portions of each of the vehicle's four propellers. Not only was the effectiveness of each propeller reduced, the drone became unbalanced and began to vibrate badly. This experiment has a myriad of applications. The most interesting is perhaps in defense scenarios where quadrotors are targeted with directed energy weapons or artillery fire. Once the dynamics of the vehicle change, non-adaptive control techniques tend to fail. Since the proposed technique is data-driven and model-free, it is not affected by the damage. Without re-tuning the controller, the performance shown in Figure \ref{fig:NoProps} was achieved. The mean absolute error for each degree of freedom was under 20 degrees per second. The rate measurements here seem very noisy; however, this is reasonable given the unbalanced nature of the injured vehicle. The authors refer the readers to the provided YouTube video to observe the smooth performance generated by the controller. An image of the vehicle with its damaged rotors is shown in Figure \ref{fig:Damaged}.

\subsection{A new vehicle: Tarot 650 Sport Quadrotor}
The controller, without being re-tuned, was transferred to a Tarot 650 Sport Quadrotor. This vehicle was equipped with high-efficiency brushless motors, heavy carbon-fiber propellers, a long-endurance flight battery, and retractable landing gear. Its center of gravity was much lower than that of the DJI. During this flight, one of the pieces of landing gear retracted upon takeoff, while the other was obstructed and remained in its initial position. This added to the complexity of the vehicle's behavior as it was unbalanced in this configuration. The proposed control technique responded well to both the vehicle swap and the unbalanced dynamics of the Tarot Quadrotor. The potential real-world applications of this experiment are again numerous as robotics rarely behave as they are designed to. The results of this flight are shown in Figure \ref{fig:Tarot}. Each of the three flights reported here are replicated in the aforementioned footage.

\subsection{Comparison to PID Control}
\begin{figure*}
    \centering
    \includegraphics[scale=1.20, trim={0.75cm 0.6cm 1cm 0cm},clip]{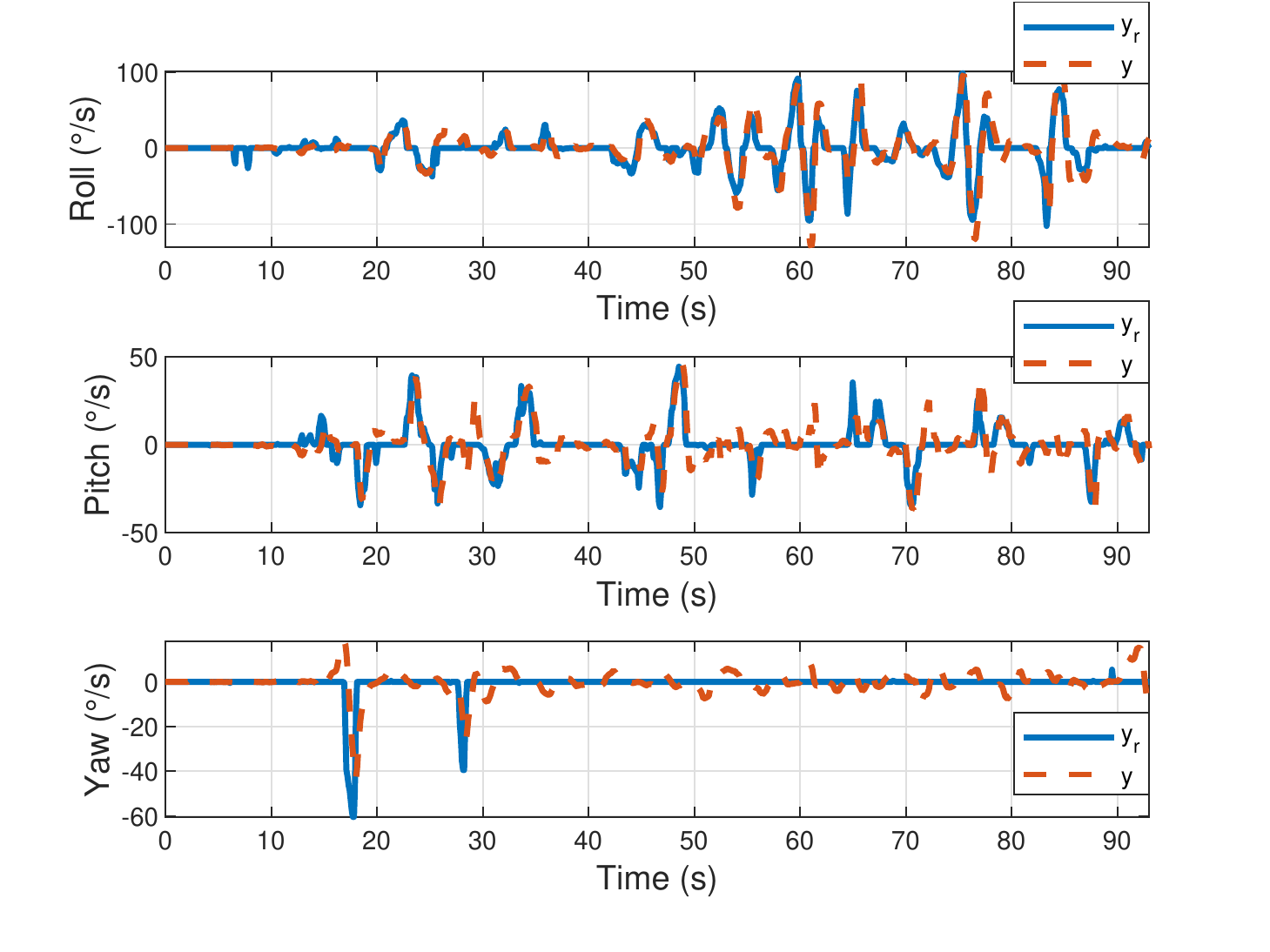}
    \caption{Tarot roll, pitch, and yaw performance under PID control.}
    \label{fig:PIDTarot}
\end{figure*}

\begin{figure*}
    \centering
    \includegraphics[scale=1.20, trim={0.75cm 0.6cm 1cm 0.2cm},clip]{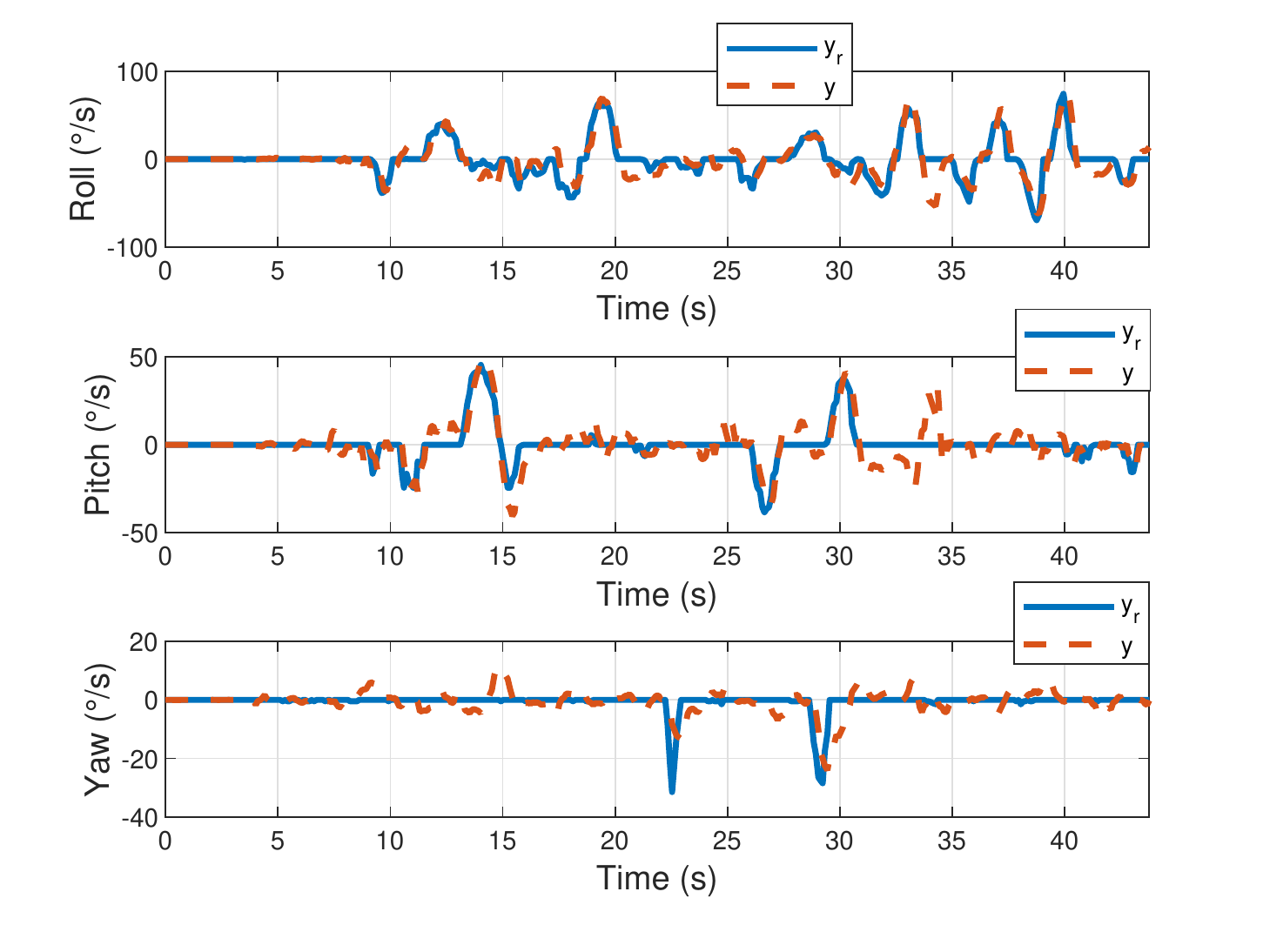}
    \caption{F450 roll, pitch, and yaw performance under PID control. No re-tuning was conducted after moving the controller from the Tarot.}
    \label{fig:PIDF450}
\end{figure*}

In order to compare the robustness of MFC to PID, a PID controller was properly tuned and deployed on the large Tarot quadrotor. The results of this flight are shown in Figure \ref{fig:PIDTarot}. After a successful flight, the controller was transferred to the DJI F450 and flown without re-tuning. The results of this flight are shown in Figure \ref{fig:PIDF450}. It is clear that MFC is much more robust across vehicles with varying dynamics. This might also suggest the ease of tuning associated with MFC as compared to PID. Note that performance of PID suggests, also, that MFC is a better multi-variable control scheme. This is particularly apparent under PID control when performing robust maneuvers on the roll axis, which seems to degrade the performance of the pitch and yaw controllers.

\section{Conclusions \& Future Work}
A highly promising control technique for quadrotors has been derived without the need of any mathematical modeling in the usual sense.\footnote{See \cite{ftc} for a connection with Machine Learning.} Even as the dynamics of the vehicle changed---either from having damage inflicted upon it or from changing the vehicle altogether---the proposed technique performed well. The little computational power required to implement this algorithm means that it could be utilized on micro and macro aerial vehicles alike. Our future work is to demonstrate a full-scale Model-Free Control architecture on a 6-DOF quadrotor---the three additional DOF being longitude, latitude, and altitude, which will come from equipping our fleet of UAVs with global positioning systems and barometers. The control scheme will look very much like the one outlined here.



\begin{figure*}[t]
    \centering
    \includegraphics[scale=0.6]{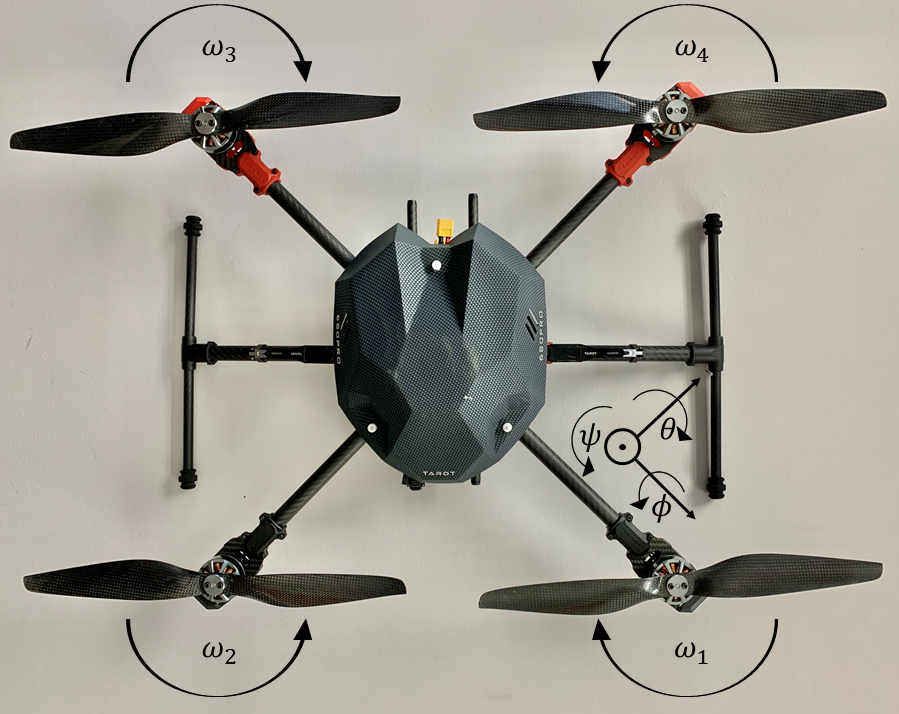}
    \caption{A Tarot 650 Sport quadrotor. This vehicle is flown in the experiments discussed in Section IV.}
    \label{fig:Quad}
\end{figure*}
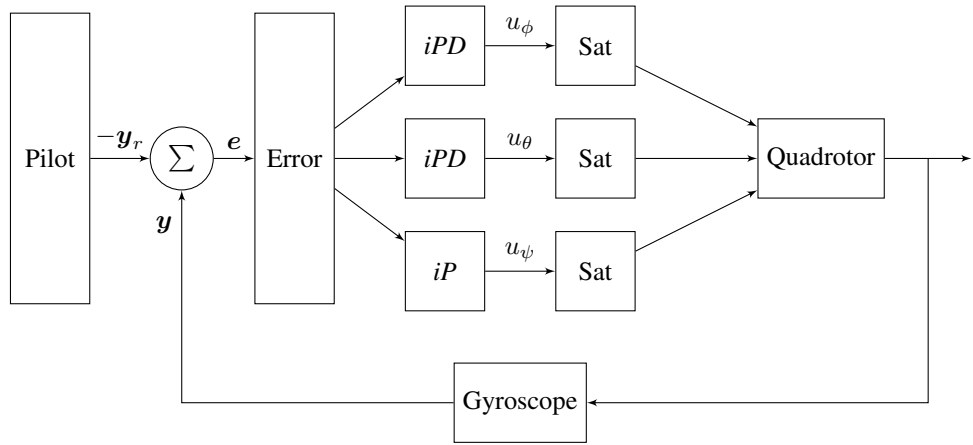
\begin{figure*}[t]
\centering
\begin{tikzpicture}[auto, node distance=2cm,>=latex']
    \node [tallblock, name=Error] (Error) {Error};
    \node [sum, left of=Error, node distance = 1.5cm] (Sum){$\sum$};
    \node [tallblock, left of=Sum, node distance = 1.75cm] (Pilot) {Pilot};
    \node [block, right of=Error] (Pitch) {\textit{iPD}};
    \node [block, right of=Pitch, node distance = 2cm](SatPitch){Sat};
    \node [block, above of=Pitch,node distance=1.5cm] (Roll){\textit{iPD}};
    \node [block, right of=Roll, node distance = 2cm](SatRoll){Sat};
    \node [block, below of=Pitch,node distance=1.5cm] (Yaw) {\textit{iP}};
    \node [block, right of=Yaw, node distance = 2cm](SatYaw){Sat};
    \node [block, right of=SatPitch,node distance=3cm] (system)
{Quadrotor};
    \node [output, right of=system, node distance=2cm] (output) {};
    \node [tmp, below of=Pitch, node distance = 3cm] (tmp1){$H(s)$};
    \draw [->] (Error) -- (Pitch);
    \draw [->] (Pitch) -- node{$u_\theta$}(SatPitch);
    \draw [->] (SatPitch) -- (system);
    \draw [->] (system) -- node [name=y] {}(output);
    \draw [->] (Error) -- (Yaw);
    \draw [->] (Yaw) -- node(Upsi){$u_\psi$}(SatYaw);
    \node [block, below of =Upsi](Gyro){Gyroscope};
    \draw [->] (Error) -- (Roll);
    \draw [->] (Roll) -- node{$u_\phi$}(SatRoll);
    \draw [->] (y) |- (Gyro);
    \draw [->] (Gyro) -| node[pos=0.92]{$\boldsymbol{y}$}(Sum);
    \draw [->] (Pilot) -- node[pos=0.5] {$-\boldsymbol{y}_r$}(Sum);
    \draw [->] (SatRoll) -- (system);
    \draw [->] (SatYaw) -- (system);
    \draw [->] (Sum) -- node{$\boldsymbol{e}$}(Error);
    \end{tikzpicture}
    \caption{Model-Free Control Scheme. ``Sat'' refers to passing the output of each control signal through a saturation function such that only reasonable commands are given to the motors of the quadrotor. The block labelled ``Error'' represents de-multiplexing the error signal and sending error values to the appropriate controller.}
    \label{fig:Scheme}
\end{figure*}


\begin{figure*}
    \centering
    \includegraphics[scale=1.20, trim={0.75cm 0.6cm 1cm 0.2cm},clip]{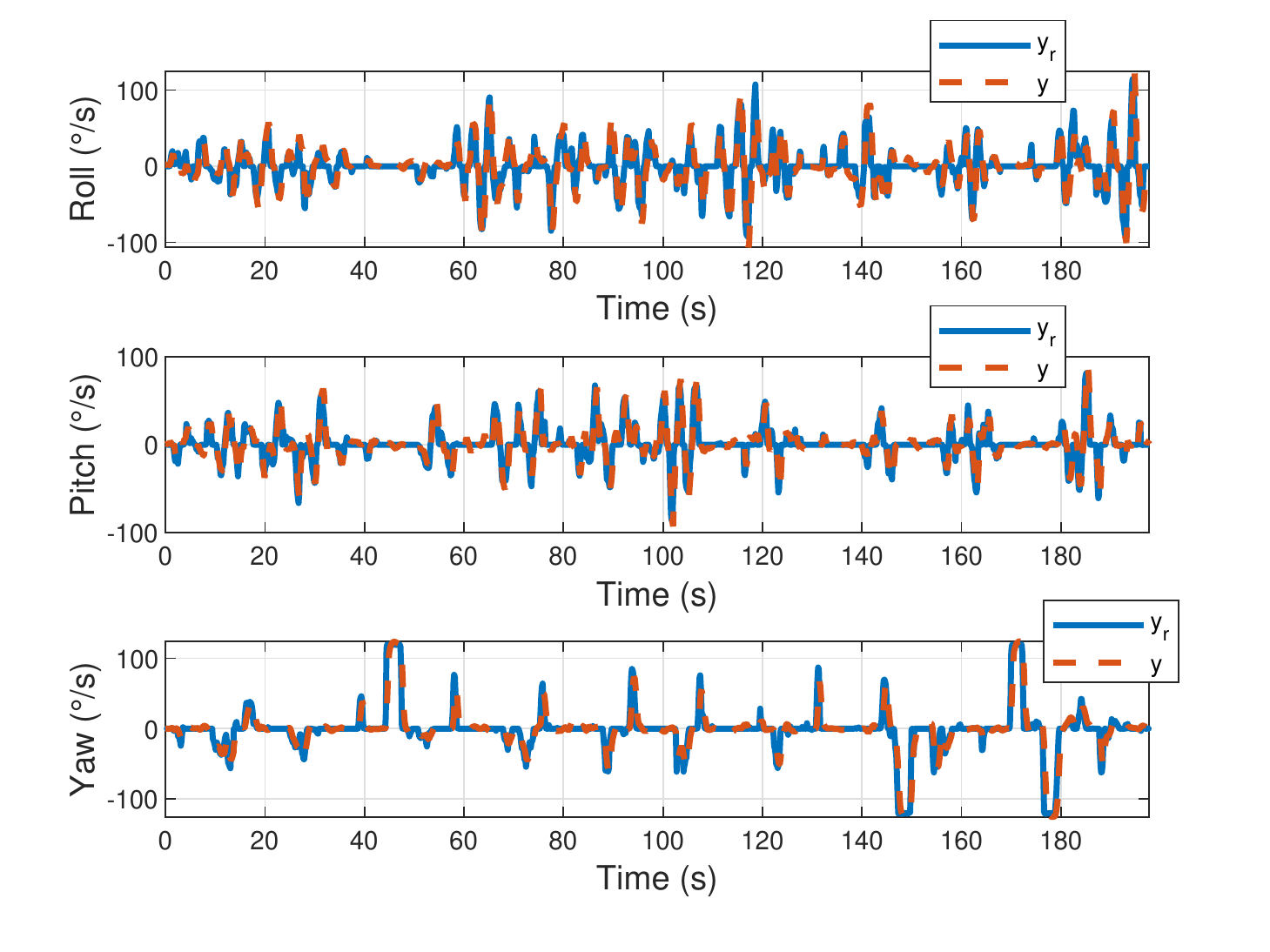}
    \caption{Tarot roll, pitch, and yaw performance after failed landing gear deployment under MFC.}
    \label{fig:Tarot}
\end{figure*}

\begin{figure*}
    \centering
    \includegraphics[scale=1.20, trim={0.75cm 0.6cm 1cm 0.2cm},clip]{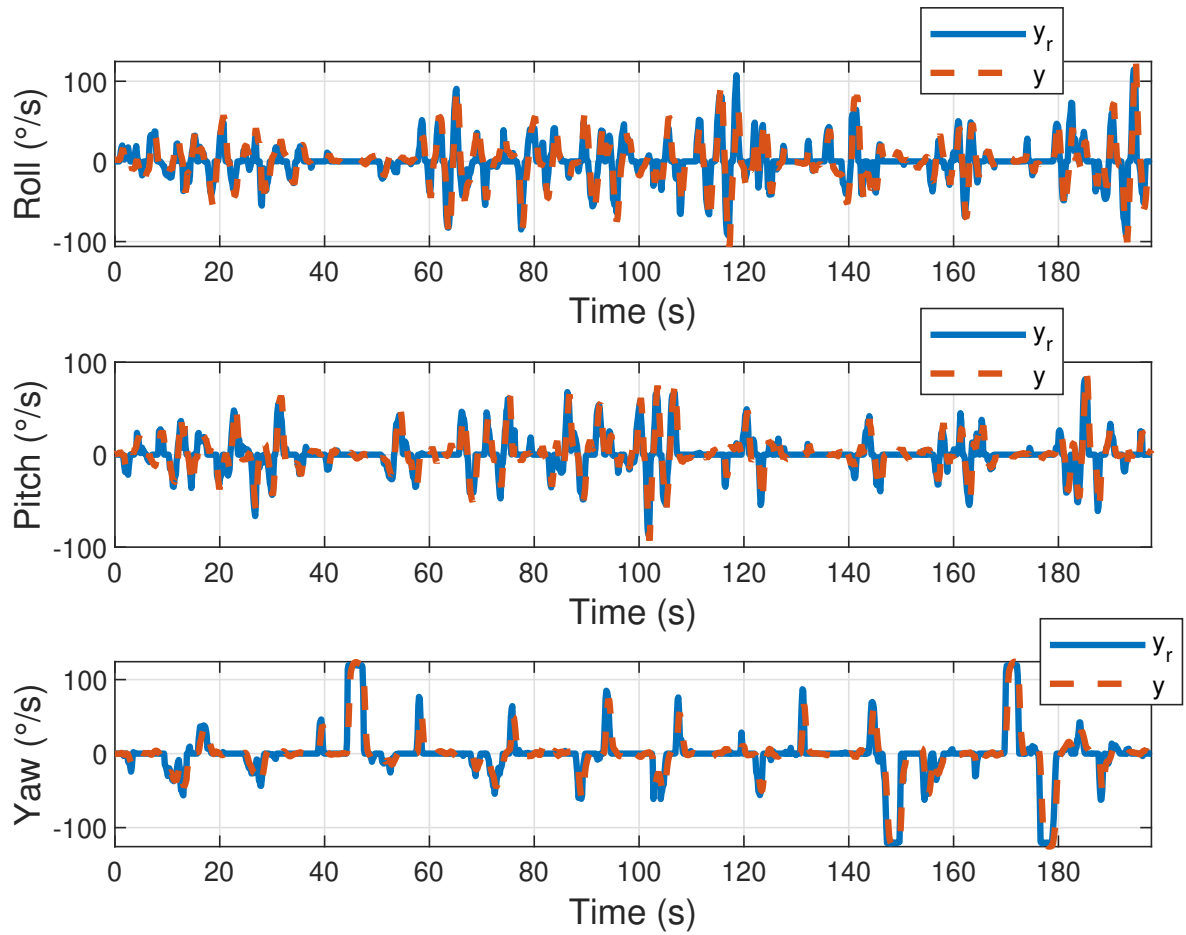}
    \caption{Tarot roll, pitch, and yaw performance after failed landing gear deployment under MFC.}
    \label{fig:Tarot}
\end{figure*}

\end{document}